\documentclass[12pt]{article}
\usepackage{graphicx,amssymb}
\begin{document}

\begin{center}
\LARGE
\textbf{Against `Realism'} \\[1cm]
\large
Travis Norsen \\[0.5cm]
\normalsize
Marlboro College \\ 
 Marlboro, VT  05344 \\ norsen@marlboro.edu \\
\end{center}

\bigskip
\bigskip

\begin{quote}
We examine the prevalent use of the phrase ``local realism'' in the
context of Bell's Theorem and associated experiments, with a focus on
the question:  what exactly is the `realism' in `local realism' 
supposed to mean?  Carefully surveying several possible
meanings, we argue that all of them are flawed in one way or
another as attempts to point out a second premise (in addition to
locality) on which the Bell inequalities rest, and (hence) which might
be rejected in the face of empirical data violating the inequalities.  
We thus suggest that the phrase `local realism' 
should be banned from future discussions of these issues, and urge
physicists to revisit the foundational questions behind Bell's Theorem.
\end{quote}

\medskip


\medskip

\section{INTRODUCTION}
\label{sec1}

I should begin by clarifying the title.  I am
actually \emph{not} against realism.  I am a realist -- at
least in several widely-used senses of the term.  What
I am against is the use of the \emph{word} `realism' in a certain
context, just as J.S. Bell was (without in any way being
professionally or morally opposed to the taking of measurements)
``Against `Measurement'.''  \cite[pages 213-231]{bell}  

The context in which I am against the use of the word `realism' 
is:  Bell's Theorem, the EPR argument, Aspect's and other
empirical tests of Bell's
inequalities, and surrounding issues.  The reason I am against
the word `realism' is twofold:  first, it is almost never clear what
exactly a given user means by the term, i.e., which of several
possible (and very different) senses of `realism' is being referred
to; and second, the point that will occupy us for most of the present
paper, \emph{none} of these possibly-meant senses of `realism' turn out
to have the kind of relevance that the users seem to think they 
have.

As far as I know, the `realism' problem was first pointed out about
ten years ago, in an essay by Tim Maudlin.  After noting, and 
answering, the long-standing misconception that Bell's theorem 
applied only to local \emph{deterministic} theories -- a misconception
Bell himself struggled against\footnote{For example, 
  in ``Bertlmann's socks and the nature of
  reality'' \cite[pages 139-158]{bell} Bell writes: ``It is remarkably
  difficult to get this point across, that determinism is not a
  \emph{presupposition} of the analysis.  There is a widespread and
  erroneous conviction that for Einstein determinism was always
  \emph{the} sacred principle.''  And there is a footnote, following
  the word ``Einstein'' which reads as follows:  ``And his followers
  [by which Bell clearly means himself].  My own first paper on this
  subject ... starts with a summary of the EPR argument \emph{from
    locality to} deterministic hidden variables.  But the commentators
  have almost universally reported that it begins with deterministic
  hidden variables.''} 
for decades, and which continues to this day -- Maudlin notes
\begin{quote}
``Recently, a new bogeyman seems to have been found: realism.
Thus Hardy states: `In 1965 Bell demonstrated that quantum mechanics
is not a local realistic theory.  He did this by deriving a set of
inequalities and then showing that these inequalities are violated by
quantum mechanics.' ....  The conversational implication
is that Bell's theorem only applies to local \emph{realistic}
theories, so that locality (and hence perhaps also consistency with
Relativity) can be recovered if one only jettisons realism.''  
\cite[page 304]{maudlin}  
\end{quote}
But, as Maudlin goes on to briefly explain, this conversational 
implication is false.

The problem only seems to have gotten worse since Maudlin's paper.
For example,
hits for the phrase ``local realism'' in the journals published by the
American Physical Society show an almost perfect exponential increase
in the last 20 years.\footnote{For 
  \emph{Physical Review Letters} alone, the number of
  papers using the phrase `local realism' for the years 1985 - 2005
  are as follows:  0, 0, 0, 1, 1, 0, 2, 5, 2, 1, 0, 0, 4, 3, 7, 4, 6,
  10, 4, 16, 13.  Note also that the rate of increase for ``local
  realism'' is significantly higher than that for other related
  keywords such as ``Bell's Theorem'' and ``hidden variables''.  So
  the increased usage of ``local realism'' cannot be blamed simply on
  the overall increase in numbers of \emph{PRL} papers generally, or papers
  pertaining broadly to the foundations of QM.}

To hint at the pervasiveness of this terminology -- and to give a
sense of how it is typically used -- here is a selection of 
statements, all from prestigious physicists
and published in peer-reviewed journals, in which the
phrase ``local realism'' (or its equivalent) appears:

\begin{itemize}

\item  ``John Bell showed that the quantum
  predictions for entanglement are in conflict with local realism.''
  \cite{zeilinger} 

\item  ``I ... illustrate the basic mathematical
conflict between the kind of predictions made by quantum mechanics and
those that Bell showed to follow from the plausible constraints of a
local realism.''  \cite{price}

\item `` `Bell's Theorem'
is the collective name for a family of arguments .... [having] the
format $E \& H \rightarrow I$ where $E$ is a description of a type of
experimental setup involving pairs of particles emitted from a common
source, $H$ is a physical hypothesis which typically expresses some
version of `realism' and some version of `locality', and $I$ is an
inequality concerning correlations....
Insofar as $H$ is typically in part a metaphysical hypothesis 
(e.g., by expressing some version of physical realism), one has 
brought experiment to bear upon a metaphysical
question.'' \cite{shimony}

\item  ``In 1964 John S. Bell .... showed
  that the tenets of local realistic theories impose a limit on the
  extent of correlation that can be expected when different spin
  components are measured. The limit is expressed in the ... Bell
  inequality.''  \cite{despagnat}

\item ``Starting in 1965, Bell and others
  constructed mathematical inequalities whereby experimental tests
  could distinguish between quantum mechanics and local realistic
  theories.  Many experiments have since been done that are consistent
  with quantum mechanics and inconsistent with local realism.'' \cite{rowe}

\item ``[L]ocal realism holds that one can
  assign a definite value to the result of an impending measurement of
  any component of the spin of either of the two correlated particles,
  whether or not that measurement is actually performed.  ....  In
  1964, however, Bell showed that ... this escape [i.e., local
  realism] from the conundrum [the EPR argument]
  is not only incompatible with the
  orthodox interpretation of quantum mechanics, but it is also
  inconsistent with the \emph{quantitative numerical predictions} of
  quantum mechanics.'' \cite{mermin1980}

\item ``Bell's
  theorem establishes that the quantum theory and the theory of
  relativity, or more properly the absence of instantaneous action at
  a distance, cannot both be correct if we wish to maintain the
  philosophical principle known as `realism.'  The absence of actions
  at a distance has come to be known as `locality,' and so Bell's
  theorem shows an incompatibility between local realism and quantum
  mechanics.'' \cite{ferrero}

\item ``Bell's theorem changed the nature of
  the [Bohr-Einstein] debate.  In a simple and illuminating paper,
  Bell proved that Einstein's point of view (local realism) leads to
  algebraic predictions (the celebrated Bell's inequality) that are
  contradicted by the quantum-mechanical predictions.... The issue was
  no longer a matter of taste, or epistemological position: it was a
  quantitative question that could be answered
  experimentally...''\cite{aspect}

\end{itemize}
And finally, Wikipedia -- that great barometer of popular
understanding (and misunderstanding)
 -- asserts the meaning of Bell's Theorem bluntly:
``either quantum mechanics or local realism is wrong.''\cite{wikipedia}

Since (roughly) the late 1970's, the claim that Bell's inequality
is a constraint on `local realism' has clearly been widespread. 
(Previously, it had been typically characterized as a constraint on local
deterministic theories or local hidden-variable theories.)
So, if my thesis (that `realism' has no valid place whatsoever in
these discussions) is correct, it follows that the underlying 
confusions are quite serious.

Of course, whether users of the phrase `local realism' 
are misusing and/or abusing the term `realism' can only be
established if we know (which, by the way, requires that \emph{they}
know) what they mean by it.  Since, unfortunately, they typically
don't tell us what they mean,\footnote{One notable exception 
is Ref. \cite{ferrero},
  which contains an admiraby detailed and at times clear-headed
  discussion of the meaning of this phrase.  Unfortunately, what the authors
  define as `realism' already includes (most of) Bell's definition
  of `locality', and what they define as `locality' turns out to
  be a pointless irrelevancy.  (Specifically, their `contextual
  variables' $\mu$ -- hidden variables pertaining to the measuring
  instruments -- could simply be shuffled into the already-existing
  symbol $\lambda$ with no loss of generality; in fact,
  this simplification would allow a shortcut to their Equation (6)
  but with the probability factors under the integrand defined more
  simply.) So their discussion is not after all a model of clarity.}
we will survey four different senses of realism that one might 
plausibly think could be relevant.

Our goal is thus to attempt to answer the
question:  what, exactly, do all these physicists mean by `realism'
when they say that Bell's inequality is based on (and hence experiment
has refuted) `local realism'?  As we will argue, no sensible answer 
is forthcoming, and so we will end with some speculation about the
origins of, and misconceptions underlying, this confused terminology.

I should clarify one other thing before we get started with our tour
of realisms.  My title is, obviously, in homage to Bell, who wrote
so eloquently (as noted above) ``Against `Measurement'.''  But the
similarity is rather limited.  Here is what Bell said about the word
whose abuse he discussed in that article:
\begin{quote}
``I am convinced that the word `measurement' has now
been so abused that the field would be significantly advanced by
banning its use altogether, in favour for example of the word
`experiment'.''  \cite[page 166]{bell}
\end{quote}
My complaint against `realism' is different in two ways.  First, I do
not think the word should be banned altogether.  As I said, I'm
actually \emph{for} several different kinds of realism, and I think
the word `realism' (appropriately specified) is perfectly good
terminology for those views, and should be kept.  What I'm against is
specifically the use of the term `realism' in the phrase `local realism' 
in the context of the EPR-Bell
issues where, I will argue, it has no place.  So my purpose isn't, 
after all, to argue that the term should be banned, but simply to 
explain why the term has no valid place in discussion of these 
particular issues.  A second (related) difference with Bell's
complaint against `measurement' is that I have no different term 
in mind (paralleling `experiment' for Bell), whose use I will urge
in place of `realism'.  Since, as I will argue, `realism' simply 
doesn't belong in these discussions to begin with, there is no need to 
replace it with some other, less misleading terminology, once its 
inappropriate use is ceased. 

Let us then begin our tour of realisms.  We will start with `naive
realism' -- or, more precisely, a certain sort of hidden variable
view that nicely parallels, in the context of physical 
measurements,\footnote{...or perhaps I should say ``physical 
\emph{experiments}''...}
the naive realist view of perception.  We will then briefly touch on
so-called `scientific realism' before moving on to `perceptual
realism' and then, finally, `metaphysical realism'.

\section{NAIVE REALISM}
\label{secNR}

In the philosophy of perception, Naive Realism is the view that all
features of a perceptual experience have their origin in some identical
corresponding feature of the perceived object.  For example, a Naive
Realist will say that, when someone sees a red apple, the experienced
redness resides in the apple, as a kind of intrinsic property that is
passively revealed in the perceptual experience.  Likewise in a case 
of experiencing the coolness of water when one plunges one's arm into
it: the Naive Realist explains the experience by positing an intrinsic
``coolness property'' of the water which is passively revealed in the
act of perception.  Naive Realism may be contrasted with
alternative theories of perception in which some aspects of
either the content or the
form of the experience is contributed, not by the perceived object,
but by the perceiving subject.  Examples of such alternatives
would include Locke's theory and the associated distinction between
primary and secondary qualities (with Naive Realism retained for the
primary qualities only), J.J. Gibson's ecological-realist 
account (according to which active interactions between the perceived
object and the subject's perceptual apparatus determine the 
form in which certain real features of the object are experienced
\cite{gibson}), 
and subjectivist accounts (in which none of the features of
the perceptual experience arise from external facts about the
object).  What all of these have in common 
is the idea that the perceiving subject (or more 
specifically his perceptual apparatus) contributes \emph{something} to
the conscious experience.  Naive Realism, in contrast to all of these,
is the belief that the identity of the perceptual apparatus
contributes nothing to the experienced product:  the experience is
simply a revealing (or passive re-creation)
of intrinsic features of the perceived object.

The philosophy of perception, however, is not our topic.  What does
any of the above have to do with physics in general or Bell's Theorem
in particular?  Quite a bit, as it turns out.  For there is a
surprisingly exact parallel between the just-described theories of
perception, and several possible attitudes toward ``measurement'' in
physics.  

For the remainder of the current paper, we will use the term Naive
Realism to refer to the following view:  whenever an experimental
physicist performs a ``measurement'' of some property of some
physical system (e.g., the position of an electron, or the 
temperature of a certain sample of liquid) the \emph{outcome} of that
measurement is simply a passive revealing of some pre-existing
intrinsic property of the object.  Thus, if the digital thermometer
reads $42.6 \; ^\circ C$, that is because, prior to the insertion of
the thermometer, the liquid already possessed the property of having
temperature $42.6 \;
^\circ C$.  And likewise, if the position measurement on the 
electron results in the electron being found \emph{here}, that is
because, prior to the measurement (i.e., prior to any interaction
between the electron and the measuring apparatus) the electron really
was, already, \emph{here}.

This last sort of case is particularly important since, according to
orthodox quantum theory, electrons don't (in general) have definite
positions -- a point made most strikingly by Bohr's colleague 
Pascual Jordan:  ``the electron is forced [by our measurement] to a
decision.  We compel it \emph{to assume a definite position};
previously it was, in general, neither here nor there; it had not yet
made its decision for a definite position.'' \cite[page 142]{bell}
According to orthodox quantum theory, the wave function alone
provides a complete description of the state of a particle, and this
wave function does not (in general) attribute any single particular
position attribute to the particle.  Thus, orthodox quantum theory
contradicts Naive Realism, and instead upholds some
physics-measurement-analogue of the non-naive realist or subjectivist 
theories of perception mentioned above.

In traditional foundations-of-physics terminology, what we are here
calling Naive Realism is thus the idea of a Non-Contextual Hidden
Variable Theory (HVT).\footnote{This 
  is to be contrasted both with orthodox quantum theory
  -- which is not a HVT in any sense -- and with Contextual HVTs such
  as Bohmian Mechanics.  Note that a Contextual HVT is a HVT for which
  (what physicists traditionally refer to as) ``measurements'' of
  \emph{at least some} properties do not simply reveal pre-existing
  values of those properties.  In Bohmian Mechanics, for example,
  position measurements \emph{do} simply reveal pre-existing particle
  positions, but everything else is contextual:  momentum, spin,
  energy, and other non-position ``observables'' do not simply have
  their pre-existing values revealed by the corresponding
  measurements.  There are several possible types of contextuality,
  all of which can be illustrated using the example of Bohm's theory.   
  In some cases (such as momentum measurements) the particle \emph{can} be
  thought of as possessing a pre-measurement value for the momentum
  (if this is simply defined as the mass times the time-derivative of
  the position), but this pre-measurement value is \emph{not}
  (generally) the
  value that appears as the outcome of a ``measurement of the
  momentum.''  (Due to the effective collapse of the wave function,
  however, the outcome of the ``momentum measurement'' \emph{does}
  match the post-measurement momentum of the particle.)  In other 
  cases (such as measurements of spin components),
  the particles don't even \emph{possess} the relevant kinds of
  properties at all:  it's not that the particle has pre-measurement
  spin components which differ, in general, from the outcomes of the
  spin component measurements; rather, the particle simply doesn't
  have any such property as ``spin'' (neither before, nor during, nor
  after the measurement).  The relevant ``spin properties'' reside
  elsewhere:  in the interaction between the particle's associated
  guiding wave and the particular measurement apparatus (Stern-Gerlach
  device, say) which is performing the measurement.  In the
  literature, the phrase ``contextual property'' is often used to
  describe features like spin as conceived in Bohmian Mechanics.  This
  terminology is unfortunate, because (as hopefully this brief
  discussion indicates) a so-called ``contextual property'' may not be
  a property at all.  See Ref. \cite{nrao} for a more detailed and highly
  illuminating discussion.  It should also be mentioned that the
  points being raised here are closely connected with Bell's reasons
  for being ``Against `Measurement'.''  The central point of his
  article is that the \emph{word} `measurement' conversationally
  implies Naive Realism -- something which, as will emerge in the
  current section of this paper, we already know with certainty cannot
  be true.  Yet the conversational implication remains, and is hard to
  resist.  Bell thought that the Naive Realist attitude that was
  implied by the misleading use of the word ``measurement'' (misleading
  because in many ``measurements'' nothing is actually being
  measured!) was behind much, if not all, of the apparent
  paradoxicalness of quantum mechanics (and the various weird attempts
  to deal with it such as ``quantum logic'').  The reader is referred
  to Bell's paper and the elaboration of Daumer \emph{et al.} for more
  details.  See also Bell's essay ``On the impossible pilot wave'' in
  Ref. \cite{bell}.}
And this raises immediately the crucial issue:  it was already known,
prior to Bell's Theorem and prior to any experimental tests of Bell's
inequalities, that Non-Contextual HVTs (i.e., Naive Realist theories)
are wrong, are not empirically viable.  It was known through the
``no-hidden-variable'' theorems of von Neumann, Gleason,
Kochen and Specker, and Bell himself.  \cite[pages 1-13, 159-168]{bell}  
(See also 
\cite{mermin2theorems} and note that the ``no-hidden-variable''
theorem of Bell referred to here is Bell's \emph{first}, less
celebrated, theorem -- the simplified version of Gleason's theorem he 
proved in the 1964 paper, infamously unpublished until 1966,
called ``On the problem of hidden variables in quantum theory''.) 
These theorems are proofs of
various versions of the following claim:  it is mathematically impossible to
consistently assign pre-measurement values to all possible observables
of a quantum system, such that (a) measurements simply reveal these
pre-measurement values and (b) the values are consistent with the
quantum mechanical predictions (for which we have strong independent
empirical support).  More specifically, the theorems show that the
value assigned to a given observable must
\emph{depend} on which (set of) compatible observables are to be
measured simultaneously -- that is, the value assigned to a given
observable must depend on the entire measurement \emph{context}, i.e.,
hidden variables must (in at least some cases) be contextual.   Or put
negatively:  Non-Contextual HVTs are ruled out.

It would be very odd, then, if the `realism' in `local realism'
meant Naive Realism, i.e., Non-Contextual hidden variables.  For then
the much-trumpeted experimental proof against `local realism' would
mean that we had either to reject locality -- for which the theory of 
relativity provides strong support -- or Naive Realism -- which is already
known to be wrong, and which we should thus \emph{already have
  rejected}.  Such a dilemma would hardly call for trumpets.
And so it seems unlikely that the `realism' in `local realism'
\emph{could} mean Naive Realism.\footnote{It should be 
mentioned, though, that this is what many of
  those who use this phrase \emph{do} appear to mean by it.  See,
  e.g., Ref. \cite{mermin1980}.  This paper is one
  of the earliest I've found (other than d'Espagnat's \emph{Scientific 
  American} article, Ref. \cite{despagnat}) which
  uses the phrase `local realism' and is notable also because the
  author actually makes some attempt to \emph{define} the phrase.
  However, two important points emerge in a footnote:  the author
  isn't himself completely clear what he means by the phrase, and 
  he confesses that his 
  `local realism' is a stronger assumption than is
  necessary to arrive at a Bell-type inequality.  This latter point
  especially will be emphasized in the subsequent discussion in the
  current paper.}

Actually, though, there is more to say here.  For the vast majority of
published proofs of Bell's Theorem do appear to assume the existence
of local, Non-Contextual hidden variables, i.e., what Mermin has
dubbed ``instruction sets'':  parameters that one thinks of as being
carried by the particles which tell them whether to be spin-up or
spin-down along various axes if the corresponding measurement is
made. \cite{nrbellproofs}   Such theories clearly exemplify Naive
Realism.  So perhaps,
after all, the derivation of Bell-type inequalities \emph{does}
require a Naive Realist premise -- and the `realism' in `local
realism' \emph{is} Naive Realism?

If this is what the users of `local realism' have in mind, however,
it simply shows that they have not properly understood Bell's
derivation.  There are two related crucial points here.

First, while it is true that many derivations of Bell inequalities use
Naive Realist ``instruction sets'', this is not an \emph{independent}
assumption.  As Bell himself repeatedly stresses (see, for example,
the remarks quoted in footnote 1 above), the existence of
these local, deterministic, non-contextual hidden variables (i.e., the
existence of Mermin-type ``instruction sets'') is not simply assumed,
but is \emph{inferred} -- from Locality plus a certain subset of the
quantum mechanical predictions, using (in essence) the EPR argument. 
\cite{nonlocchar}  Logically, therefore, it is misleading
to claim that the Bell inequalities are derived from 
Locality \emph{and} Naive Realism -- and hence equally misleading
to claim that empirical violations of Bell inequalities
permit some kind of choice between rejecting Locality and rejecting
Naive Realism.  Any such choice is illusory, for the (modified) EPR
argument proves that Locality \emph{entails} Naive Realism.  Thus, to 
have to choose one of them to reject, is to have to reject Locality.

Second, it is possible to derive a Bell-type inequality \emph{without}
the Naive Realist instruction sets (i.e., without the EPR-motivated 
deterministic, non-contextual hidden variables).  In particular, the
so-called CHSH inequality (named for it discoverers Clauser, Horne, 
Shimony and Holt \cite{chsh}) can be derived from the assumption of Locality
\emph{alone}.  Nothing else (such as determinism, hidden variables,
counter-factual definiteness, etc.) need be assumed. \cite{cfm} 
This fact,
unfortunately, was obscured in the original CHSH paper.  They
described their own derivation as being based on the 
assumed existence of hidden variables which go beyond the orthodox
quantum description of states in terms of wave-functions only:
``Suppose now that [the outcomes $A$ and $B$ are] 
due to information carried by and localized within each
particle... The information, which emphatically is not quantum
mechanical, is part of the content of a set of hidden variables, denoted
collectively by $\lambda$.''  Despite these statements, however, the
mathematical derivation itself requires no assumptions whatever about
the content of $\lambda$.  In the context of Bell's definition of
local causality (i.e., Bell Locality), $\lambda$ refers to a theory's
proposed \emph{complete description} of the state of the particle pair
prior to measurement.  But -- and this generality is precisely why
Bell's theorem is so \emph{interesting} -- this could be \emph{any}
theory.  For example, orthodox quantum theory is in principle covered,
with the identification $\lambda = \Psi$.  Thus, the only assumption
actually used in the formal derivation of the CHSH inequality is
(Bell's ``local causality'', i.e., Bell) Locality.  And so the
empirically observed violation of the CHSH inequality can only be
blamed on a failure of the Locality assumption, i.e., the nonlocality
of nature.  

To summarize, one simply does not need an assumption of Naive Realism
in order to derive an empirically testable Bell-type inequality.  One
needs only the Locality assumption.  And so, whatever the `realism'
in `local realism' is supposed to mean, it cannot be Naive Realism.

Before moving on to our other candidates for the meaning of
`realism' in `local realism', let's address one possible worry.
Perhaps `realism' refers to some other, less naive, class of hidden
variables (less naive, that is, than the Mermin-type
deterministic non-contextual instruction sets).  For example, perhaps
we are supposed to understand the `realism' in `local realism' as
allowing some sort of non-deterministic or contextual hidden
variable theory.  

But the two arguments already given show that this cannot be right.
Taking them in the opposite order this time, the CHSH inequality can
be derived without any hidden variable assumption at all
(non-contextual or otherwise; deterministic or non-deterministic; 
naive or sophisticated) and, anyway, an appropriately-modified EPR
argument proves that Locality \emph{requires} the naive, Mermin-style 
deterministic non-contextual instruction sets. \cite{nonlocchar}  So the
`realism' in `local realism' must not only not be Naive Realism as
we have defined it here -- it must not be any kind of hidden variable
assumption \emph{at all}.  And so if we are looking for a sense of
`realism' whose use in `local realism' would actually be
warranted, we will have to look elsewhere.

\section{SCIENTIFIC REALISM}

What other senses of `realism' might be intended by the authors who
claim that `local realism' is refuted by Bell's theorem and the
associated experiments?  Perhaps they mean Scientific Realism.  In the
philosophy of science, this is the doctrine that we can and should
accept well-established scientific theories as providing a
literally-true description of the world.  Scientific Realism is the
doctrine that we should \emph{believe} the ontologies posited by our
best scientific theories.  It is normally contrasted with
Instrumentalism -- the idea that theories are merely ``instruments''
for making empirical predictions, and their ontologies (especially in
regard to unobservables) are not to be taken literally.  
For example, because of the overwhelming
theoretical and empirical success of the atomic theory of matter, a
Scientific Realist would urge us to accept
that, in fact, matter is made of atoms -- that the ontology of atoms is not
merely a useful fiction (as the Instrumentalist would hold) but a
literally true description of matter's actual constitution. \cite{scirealism}

Perhaps the users of `local realism' think that Bell's Theorem is
likewise based on the assumed-literal-truth of some particular 
scientific theory
-- that is, perhaps what they mean by the `realism' in `local
realism' is Scientific Realism.

But this, I think, can be dispensed with immediately.  The most widely
accepted theory of the phenomena relevant to Bell's Theorem is
orthodox quantum mechanics (OQM).  
So if an appeal to Scientific Realism were
being made here, it would evidently be made in support of OQM.
But nobody (I think) believes that Bell's inequality is premised on the
assumed truth of OQM; if anything, people widely
accept the opposite -- that Bell's theorem refutes only some odd,
un-orthodox sort of theory that one probably shouldn't have believed
in anyway (and so, whatever type of theory that is, its allegedly being assumed
as a premise in the derivation could hardly be motivated by an appeal
to Scientific Realism).  

And even this represents a confusion.  For, as mentioned above,
Bell-type inequalities can be derived without any assumptions whatever
about the specific nature of the covered theories (other than that
they respect Locality).  That is, \emph{all} local theories must
respect the inequalities (and are hence apparently
ruled out by the experiments).  
Nature, therefore, is not local (in the sense of respecting relativity's
prohibition on superluminal causation).\footnote{See
  Ref. \cite{nonlocchar} for a careful discussion of Bell's definition
  of Locality, which we use throughout.}

Someone who grasps that this is the correct way to understand the
significance of Bell's theorem -- but who misunderstands the content
of the theorem -- might plausibly invoke Scientific Realism.  For
example, someone might erroneously believe that Bell's argument for
non-locality consists of the following:  OQM (or
Bohmian Mechanics, or whatever one's favorite empirically-viable
version of quantum theory happens to be) is a good, widely-accepted,
theoretically and empirically successful theory -- and it is
non-local -- so therefore nature is non-local.  Such an argument would
invoke Scientific Realism in justifying the leap from OQM (or
whichever) being a ``good, ...'' theory, to its corresponding with
nature.  But, obviously, such an argument simply represents a 
confusion.  It is precisely the fact that Bell's Theorem is
\emph{general} -- that it is \emph{not} based on the assumed truth of 
any particular candidate theory -- that makes the theorem so
interesting and so profound.

What about the assumption which \emph{is} required to arrive at a
Bell-type inequality, namely Locality?  Isn't this supposed to be
motivated by relativity theory (and specifically its prohibition on
superluminal causation), and isn't an appeal to Scientific Realism
needed to warrant taking this prohibition seriously, as a real fact
about nature?  But this wouldn't help in making sense of `local
realism', for that phrase clearly makes explicit already the Locality
assumption.  And once that assumption is made explicit, there is no
point in specifying additionally the reason why one should take the
assumption seriously.  Locality is an assumption from which Bell-type
inequalities can be derived; that it is assumed, is perfectly adequate
to make the logical structure of the argument clear.  And anyway, the
upshot of the argument is precisely that the Locality assumption
cannot be correct.  No matter how seriously one takes it, experiment
refutes it.  So while Scientific Realism might have some role to play
in emphasizing the profundity of Bell's theorem,\footnote{Or more
  likely, Bell's theorem would be used as an argument against some
  versions of Scientific Realism (namely, versions which take
  something like ``the concensus of the scientific community'' as the
  required warrant for believing in the literal truth of a theory).
  For Bell's theorem and the associated experiments prove that --
  the ``concensus of the community'' for most of the last century to 
  the contrary notwithstanding --
  some important aspects of relativity theory are either flat wrong,
  or less fundamental or universal than nearly everyone believed.}
it cannot help us in understanding the meaning of `local realism'.

So, it seems, the `realism' in
`local realism' cannot mean Scientific Realism.  We will have to dig
deeper if we are to find some justification for this popular
terminology.

\section{PERCEPTUAL REALISM}

By ``Perceptual Realism''\footnote{The 
  arguments in this section are similar to, and inspired
  by, Tim Maudlin's discussion in Ref. \cite{maudlin}  
  Note that what I call Perceptual Realism is related to the
  property of theories that, in Maudlin's terminology, makes them
  ``standard''.  Specifically, the relation is this:  any of Maudlin's
  ``standard theories'' can be accepted as true without having to deny
  the truth of ordinary perceptual judgments, i.e., without rejecting
  Perceptual Realism.  As Maudlin explains, ``Any non-standard
  interpretation must do considerable violence to our basic conception
  of the physical world.  ...[I]t cannot contain physical events or
  objects ... that even vaguely correspond to the world as it appears
  to us.  If [for example] it seems to us that a needle on a piece of
  apparatus moves one way rather than another at a particular time and
  place, or a particle detector clicks at a particular time and place,
  a non-standard interpretation must insist that nothing at all which
  reflects those supposed events takes place in the corresponding
  regions of space-time.''}
I will mean the idea that sense perception
provides a primary and direct access to facts about the world -- i.e.,
that
what we are aware of in normal perception is \emph{the world}, and not
any sort of subjective fantasy, inner theater, or mental construction.  
\cite{kelley}  Note
that this presupposes that there \emph{is} an external world -- a
doctrine I will call Metaphysical Realism and to which we will turn
later.  Perceptual Realism is specifically the claim that ordinary
sense perception provides valid information \emph{about} this 
external world, that
the appropriate perceptual experience provides a sufficient basis for
accepting the \emph{truth} of ordinary perceptual judgments (e.g., 
``There is a table in front of me'', ``My cat Buster is looking out 
the window'', and ``That light is currently glowing red, not
green'').  


Perceptual Realism denies that we are
systematically \emph{deluded}, about the real state of the world, by 
our perceptual experience.  It may thus be contrasted, for example, to
Platonic Idealism, according to which true reality is 
nothing like the familiar perceptual world of material objects moving
and interacting.  If, really, there is \emph{not} a table in front of
me (or there both is and isn't a table, or there is really no such
thing as solid entities such as tables, or I am really a brain in a
vat and the whole of my perceptual experience is a delusion fed to me
by evil scientists) then Perceptual Realism would be false.

It is worth noting at the outset that Perceptual Realism is the
foundation of \emph{empiricism} and hence of modern empirical
science.  Leaving aside the possibility of cognitive nihilism, any
denial of Perceptual Realism will necessarily put forth some alleged
alternative to sense experience as the proper source of ideas, i.e.,
as our primary means of contact with the external world.  Such
alternatives (e.g., mystic revelations, rationalistic deductions from
a priori self-evidencies, innate ideas, instincts and intuitions)
are familiar to most scientists -- familiar, that is, as the kinds of
nonsense we have to fight against \emph{as} scientists.  

At the risk
of giving a false impression of the narrowness of Perceptual Realism,
let us briefly underline its importance to empirical science by
pointing to an important class of examples.  The meaning of
``empirical'' in ``empirical science'' is the idea that our more
abstract ideas (e.g., scientific theories) are grounded ultimately in
data about the world that comes from experience -- and specifically,
for theories, \emph{experiment}.  And what is
an experiment?  It is a controlled intervention in nature from which
we may infer something about nature -- in short, experiment is our 
way of posing specific pointed questions to nature, and receiving
equally pointed answers.  Note especially the obvious importance of
our being able to receive the answer.  In a typical experiment, the
\emph{outcome} will be registered in the position (or some other
perceptually-obvious feature) of a macroscopic object -- e.g., the
position, along some scale, of a pointer or ``needle'', the color of
some flashing light, or a number projected on a computer screen or
printed on a sheet of paper.  What we wish to stress here is the role
of Perceptual Realism in grounding our ability to become aware of the
outcome of the measurement.  If we can look at the flashing light and
be systematically deluded about its color (e.g., we think it's red
when it's green and vice versa, or our seeing red or green has no
correlation whatever to the actually-flashed color) then Perceptual
Realism is false -- and empirical science is hopelessly doomed.  

What could such a fundamental philosophical principle as Perceptual
Realism have to do with Bell's Theorem?  Let us get into this by
raising the example of the Many Worlds Interpretation (MWI) of quantum
theory, which has occasionally been suggested as a counterexample to
the understanding of Bell's theorem I have expressed above -- namely,
that Bell's theorem (and the associated experiments) prove that no 
local theory can be empirically viable.  Orthodox quantum mechanics 
is certainly not a
counterexample to this claim: it is manifestly nonlocal, with the
nonlocal dynamics appearing specifically in the so-called ``collapse
postulate''.  The basic motivation of the MWI is to restore Locality
by simply dismissing, as unnecessary, the collapse postulate and
retaining only the (local) unitary dynamics (Schr\"odinger's equation 
or its appropriate generalization).  The resulting theory is then
manifestly local, thus (it is claimed) proving that, after all,
relativistic causality can be maintained in the face of Bell's
Theorem. 

Of course, this program runs quickly into the problem of
Schr\"odinger's cat.  Without the collapse postulate, the cat does not
end up dead \emph{or} alive, but, with certainty, in an entangled
superposition of dead \emph{and}
alive.  Specifically, the unitary dynamics generates a massively
entangled state in which the cat (and likewise virtually all familiar
macroscopic objects) do not possess the familiar sorts of determinate
properties (such as being definitely alive or definitely dead,
definitely here as opposed to definitely there, etc.).  In other
words, the ontology posited by MWI -- the story it tells about the
actual state of the external world -- is radically at odds with our
perceptual experience.  We \emph{see} a cat that is either definitely
alive or definitely dead, but according to MWI these perceptions are
\emph{delusional}.  Really -- in actual fact -- the cat is \emph{not}
definitely dead, and it is \emph{not} definitely alive either; it is
in an (entangled) superposition of dead \emph{and} alive.  And, simply
put, that is not what we see when we look.  Different versions of the
basic MWI theory 
\cite{albertmwi}
give different accounts of the precise relation
between consciousness and the posited real state of the world -- but what they
all have in common is a radical failure of correspondence between
perceptual experience and this posited real state, i.e., what they have in
common is the need to reject Perceptual Realism.

The crucial point is that MWI is, in fact, not a counterexample to the
understanding of Bell's Theorem I've advocated here -- namely, that
what Bell's Theorem proves is that no local theory can be
consistent with the data acquired in (for example) Aspect's
experiment.  MWI is not a counterexample to Bell's claim that no
local theory can explain that data.  Instead, it is (by implication) 
an invitation to reject that data as fallacious, to reject 
as a \emph{delusion} the belief that the experiments actually 
\emph{had} the outcomes reported in Aspect's paper.  

Let us be painfully concrete.  Imagine Alice and
Bob sitting at spatially-separated locations, randomly setting the
dial on their (say) Mermin-type ``contraptions'' (i.e., rotatable 
Stern-Gerlach devices or polarizing filters), 
noting whether the light on top of the 
device flashes red or green for a given run, and then writing this 
outcome down in a lab notebook so that their two data sets can be
later compared and the appropriate correlation coefficients computed.  
The point is:  for each run of the experiment, Alice perceives that
either the red light or the green light has flashed, and writes down
(or, more accurately, attempts to write down and erroneously believes 
herself to be writing down) the corresponding outcome in her lab notebook.  
But, according to MWI,
\emph{every single one of these reports is false}.  In actual fact,
according to MWI, what happens as each particle passes through the
measurement apparatus is that the apparatus gets into an entangled
superposition of the red-light-flashing and green-light-flashing
states.  Thus, according to MWI's description of the world, for 
\emph{none} of the experimental runs did the light flash
one or the other of the definite colors.  Alice's perception to the
contrary is a \emph{delusion}, and so her data notebook is full of
falsehoods,\footnote{Of course, one should really follow the unitary
  evolution into the pencil marks in the notebook, and say that those
  too end up in complicated entangled superpositions -- thus rendering
  the marks more plausibly consistent with the real outcomes of the
  experiments.  But this just moves the problem back without answering
  it, for Alice (and, later, Bob) \emph{sees} pencil marks
  indicating either ``red light flashed'' or ``green light flashed''
  -- and, on this account, neither of these corresponds to the real
  state of the pencil marks in the notebook.  Or we can  move it even
further:  when Alice and Bob get together later to exchange results
and compute correlation coefficients, Alice \emph{hears and sees} Bob
reporting that, for example, on run number 42 he saw a red
flash -- even though, according to MWI, in actual fact, Bob reports no
such thing.  Or further yet:  when I open the dusty old copy of
\emph{Physical Review Letters} in the library to see what Aspect
reports as the outcome of his experiment, I \emph{see} the droplets 
of ink on the pages arranged in a way that I take as meaning that
Bell's Inequality was violated in the experiment -- even though,
according to MWI, in actual fact, the droplets are \emph{not} so
arranged.  The point is, on the basis of some kind of perceptual
experience, \emph{I come to believe something 
definite about how a certain set of measurements actually came out}.
But this belief, according to MWI, does not correspond to  what in
fact happened; the belief is a delusion.}
and so the real relationships between Alice's and Bob's experiments
are not reflected in the 
correlation coefficients that end up getting
reported in the published paper.  According to MWI,
it's not that Bell's Theorem (as I have explained its implications 
earlier) is wrong; \emph{it's that we are wrong to think that Bell's
inequalities are, in fact, violated.} \cite{albert}

So might the `realism' in `local realism' mean Perceptual Realism?
The idea of MWI as a counter-example to the claim that Bell proved the
inevitability of non-locality, might have suggested this.  But as we
have argued, MWI drives its wedge not into this understanding of
Bell's Theorem per se, but, rather, into the idea that the outcomes of
certain experiments were what we thought they were (based on, among
other things, direct perception of the positions of pointers, the
colors of flashing lights, or the patterns of ink on the pages of
physics journals.  Putting the
same point differently, Perceptual Realism is \emph{not} an assumption
that goes into the derivation of Bell's inequalities, so it would make
no sense to interpret theorist's claims that the inequalities reflect
`local realism' as referring to Locality and Perceptual Realism.  

On the other hand, Perceptual Realism \emph{is} needed to arrive at
the claim that Bell's inequalities are, in fact, violated.  But this
doesn't help make sense of the phrase `local realism' either, since
an experimentalist could never claim to have empirically refuted
the `realism' in  
`local realism' if `realism' means Perceptual Realism.  It's not
that the experimental data leaves open a choice between which of two
premises -- Locality or Perceptual Realism -- to reject.  To accept
the data at face value 
\emph{is} implicitly to endorse Perceptual Realism -- thus leaving
Locality as the only possible premise to reject.  One could indeed
follow MWI in retaining Locality by rejecting Perceptual Realism,
but \emph{this is not a move one makes as a response to the experimental 
data}; rather, \emph{it is a move one makes to justify rejecting the data as
systematically failing to reflect the true state of the world.}

So it does not seem possible that the `realism' in `local realism'
means Perceptual Realism.

Since we have raised the issue, it is worth spending a moment to
assess MWI's strategy of maintaining Locality by rejecting
Perceptual Realism.  The problem with this strategy is implicit in
what we've said already about the fundamentality of Perceptual Realism
to modern empirical science, but it is worth making this more
explicit since many MWI advocates seem to underestimate the price they
are paying to save Locality.  

Consider a hypothetical example:  a new drug is discovered
which, it is thought, might have cancer-fighting properties.  So an
empirical trial is undertaken, in which cancer patients are randomly
assigned either the new drug or a placebo.  After several years, the
outcome is not good:  \emph{all} of the patients given the drug have died,
while the death rate among those given the placebo is around 50\% --
typical, let us say, for similar unmedicated patients.  

The obvious inference here is that the drug has a negative effect on
the health of the cancer patients:  it doesn't cure them, it
\emph{kills} them!  But a different conclusion could be reached if we
are willing to entertain a rejection of Perceptual Realism.  Suppose a
medical researcher proposes a theory according to which giving this
drug to cancer patients has two effects:  first, the patients are cured of
their cancer, and second, the doctors who dispensed the drugs are afflicted
with a permanent hallucinatory state in which they (delusionally)
believe that their patients have died.  Still suffering from these
delusions, they write articles for JAMA reporting the data as I
described it in the previous paragraph, and conclude that under no
circumstances should this drug (which is in fact, according to this
theory, the cure for cancer!) be given to any more cancer 
patients.\footnote{To make the story more closely parallel to MWI, we should
  add that the permanent hallucinatory state afflicts everyone else,
  too, so that the real state of the patients becomes \emph{in
    principle} unobservable.}

Does any scientist think that such a ``theory'' could or should
be taken seriously by the medical community?  (That is a purely
rhetorical question to which the answer is obvious.  But the following 
question probably warrants explicit and open discussion, since the
physics community apparently does not regard its answer as obvious.)  
And isn't the Many Worlds Interpretation of quantum theory precisely 
parallel to this in all relevant respects?

Advocates of MWI typically try to spin things away from the direction
I've just indicated.  It's not, they argue, that our perceptual
judgments are \emph{delusional} -- rather, it's only that they are
\emph{incomplete}.  When we see a living cat or a green light, it's
not that our experience fails to correspond to the real state of the
world -- rather, we experience only \emph{part} of the world,
specifically, one of the many ``worlds'' (or more accurately, one of
the many terms in the universal wave function which completely
describes the
real state of \emph{the} world).  So, they claim, the apparent
non-correspondence between my perceptual experience of (say) a living cat
and the real state of the world (which involves the cat being in an
entangled superposition of alive and dead), is no more problematic
than the fact that, for example, 
I can perceive (currently) the objects in this 
room but not the top of the Empire State Building.  To perceive a 
\emph{part} of the whole universe, is still to perceive validly.  One
cannot take omniscience as the standard of valid perception,
and so (it is argued) MWI actually does not require a rejection of
Perceptual Realism.

This objection, however, trades on 
a significant abuse of the word ``part''.  I accept as
a crucial principle that one must reject omniscience as the standard
of validity, across all of epistemology.
To be perceptually aware of
some fact \emph{is} to be perceptually aware, and this awareness does not
become delusional merely because there are some additional facts out
there in the world of which one is not (currently) aware.  But this 
only helps the MWI advocate answer the problems I've pointed out above
if it is correct to interpret the various terms in the (massively 
entangled) universal wave function as each,  individually,
representing a state that is somewhere realized.  For example, take
the case of Schr\"odinger's cat, and suppose the quantum state 
function for the cat's various degrees of freedom takes the form
\begin{equation}
|\Psi \! > = \frac{1}{\sqrt{2}} \left( |{\mathrm{Alive}}\!> +
 |{\mathrm{Dead}}\!> \right).
\end{equation}
The point here is that if, when the real state of the cat is 
described by $|\Psi\!>$, someone believes that the cat is definitely
alive, that person's belief is not a ``partial truth'' but a plain
ordinary falsehood.  For what it \emph{means} for the cat to be
``definitely alive'' is (according to the theory here in question)
for its degrees of freedom to be described, not by $|\Psi\!>$, but
by the quantum state $|{\mathrm{Alive}}\!>$.  And, by hypothesis,
that is simply not the state the cat is in.  The person's belief is as
wrong as it would be if they believed the cat was definitely alive
when, in fact, its quantum state was $|{\mathrm{Dead}}\!>$.  In this
regard, the states $|{\mathrm{Dead}}\!>$ and $|\Psi\!>$ are equivalent:
they are both \emph{not} the state ($|{\mathrm{Alive}}\!>$) which
would have to obtain to render the person's belief true.

Perhaps some advocates of MWI are fooled by the theory's name
(which is in fact a misnomer).  If there were some sensible  way of taking
the many individual terms in the universal wave function to represent
literally distinct universes, perhaps it could make sense to interpret
a belief like the one considered in the last paragraph to be a
``partial truth'' (since the belief would then correspond to a fact
that is realized in at least that one universe, and would hence indeed 
be true).  And then perhaps
an advocate of MWI could still consistently endorse Perceptual Realism.
But, in fact, one cannot think about the terms this way (since, among
other reasons, what distinct universes exist would then depend on our
arbitrary choice of basis states).  No, to make sense of MWI, we must
accept that there is just a single universe and that its complete
physical description is provided by the massively entangled wave
function we get from solving Schr\"odinger's equation (and never
applying the collapse postulate).  And the price of that is
unavoidably to give up the idea that our common sense (perceptually
based) beliefs correspond to the actual state of the world.  In other
words, the price is the rejection of Perceptual Realism.

And this brings us back to our earlier claim that Perceptual Realism
is a foundational principle for modern empirical science.  To
seriously entertain a scientific theory which requires us to reject
Perceptual Realism is to engage in a vicious sort of large-scale 
circularity, as David Albert has pointed out. \cite{albert2} 
To the extent that a
theory poses as scientific, it asks to be considered as a possible
best explanation of a certain class of empirical data.  In the case
of MWI, this includes primarily all of the data on which
Schr\"odinger's equation and its various relativistic extensions  
rest.  But at the same time, the associated need to reject
Perceptual Realism requires us to dismiss \emph{that same data} 
as not actually reflecting the true state of the world.  A theory like
MWI would evidently have us dismiss as delusional the very evidence
that is supposed to ground belief in the fundamental equations that
define the theory -- a very uncomfortable logical position, to be sure.  

Let us formulate this important point in positive form.  There is no 
possibility that one day in the future scientists will go into a 
laboratory, do some sophisticated experiments, and infer from the 
outcomes of those experiments that our eyes systematically delude us
about the state of things in the world.  Such a scenario is impossible
because it involves a logical contradiction:  the conclusion reached
by the imaginary future scientists undercuts the imagined
evidentiary basis for that conclusion.  The claim that the
conclusion should be believed \emph{because of that evidence}, is
therefore \emph{self-refuting}.  
Perceptual Realism is thus an \emph{axiom} (in the Aristotelian sense
of passing the test of reaffirmation-through-denial) for modern 
empirical science:  any allegedly empirical-scientific argument
against Perceptual Realism would necessarily
be self-refuting.  Looked at this
way, it is hard to understand what kind of evidence an MWI advocate might 
offer in favor of that theory.  It is simply not very convincing to 
say that the theory offers the best possible explanation  of a bunch
of events in physics labs over the last 100 years -- events which,
according to the theory, didn't actually happen.

There is one other important point to be made against MWI's being
taken seriously as a viable version of quantum theory.  For MWI, the
rejection of Perceptual Realism is general.  It requires us to reject 
not just the data apparently showing violations of Bell's
inequalities, and not just the data underlying the specific equations (e.g.,
Schr\"odinger's) that define the dynamics of that theory, but to
reject, in principle, all the data coming from all experiments.  And
this includes, in particular, all of the experimental data that is
normally taken to support relativity theory and the associated account
of space-time structure -- the saving of which was the only real
motivation for taking MWI seriously in the first place!  So not only
is MWI apparently self-refuting in terms of its actual dynamical
content; it is self-refuting also in regard to its basic motivation.  
As Maudlin explains this point, accepting MWI would mean accepting that
\begin{quote}
``physical reality contains nothing like a relativistic space-time
containing localized events and objects which even approximately
correspond to the events and objects we think we see.  In such a
circumstance, it is hard to see why we would continue to hold the
relativistic account of space-time structure seriously, since that
account is based on observations which were taken to report objects
and events in space-time.  In short, it is hard to see why we would
seriously believe that we had gotten the deep structure of space-time
right if we had gotten questions about whether, for example, a needle
on an instrument actually moved to the right or the left wrong.''
\cite[page 287]{maudlin}
\end{quote}
The bottom line is the impossibility of any scientific basis for any 
(allegedly scientific) theory requiring the rejection of Perceptual
Realism.  MWI requires such a rejection, and hence cannot be taken
seriously as a scientific theory.  But, to return to the main
development, this is only relevent by way of refuting the idea that
the `realism' in `local realism' might justifiably denote Perceptual
Realism.  As we argued earlier, it doesn't, so we will have to
continue digging if we are to find some relevant sense of
`realism'.

\section{METAPHYSICAL REALISM}

Despite the harsh criticisms of MWI in the previous section, there is
one aspect of the theory which I fully support:  it accepts the
existence of a single, objective, external world ``out there'' 
whose existence and
identity is independent of anyone's awareness (or, in the case of MWI,
non-awareness) of it.  
That is, even MWI endorses what I will call Metaphysical
Realism.  This Realism accepts the existence of an external world, but
without necessarily requiring anything specific in regard to its
similarity to the world of our perceptual experience or the account of
any particular scientific theory.  
What can Metaphysical Realism be contrasted
with?  It seems the only possible contrast would be outright
solipsism -- the doctrine that ``it's all just ideas in my head.''
Even a thoroughgoing subjectivist idealism which says (say) that we
all create our own experience out of whole cloth, evidently acknowledges the
real, objective existence of (at least) those other 
(subjective-experience-creating)
individuals.  Likewise, the traditional brain-in-vat scenario must
accept the real physical existence of brains, vats, and the evil scientists (or
computers or whatever is running things).  To reject Metaphysical
Realism one must reject the real external existence of \emph{anything}
outside of one's own mind -- i.e., one must endorse solipsism.  

The implication is that, if one is to use any words with anything like
their ordinarily intended meanings, one is tacitly assuming
Metaphysical Realism.  So it should not be surprising that Bell's
Theorem (a specific instance of, among other things, using certain
words with their ordinary meanings) rests on Metaphysical Realism.
This manifests itself most clearly in Bell's use of the symbol
$\lambda$ to refer to a (candidate theory's) \emph{complete
  description} of the state of the relevant physical system -- a usage
which obviously presupposes the real existence of the physical system
possessing some particular set of features that are supposed to be
described in the theory.
Putting it negatively, without Metaphysical Realism, there can be no
Bell's theorem.  Metaphysical Realism can (thus) be thought of as a
premise that is needed in order to arrive at a Bell-type inequality.

And so it seems we may have finally discovered the meaning of the
`realism' in `local realism'.  One \emph{cannot}, as suggested
earlier, derive a Bell-type inequality from the assumption of Locality
alone; one needs in addition this particular Realism assumption.  This
therefore explains the `local realism' terminology and explains
precisely the nature of the two assumptions we are entitled to choose 
between in the face of the empirical violations of Bell's inequality.
On this interpretation, we must either reject Locality or reject
Metaphysical Realism.  

I do not know for sure that this isn't what the users of `local
realism' have in mind.  There is, in favor of this interpretation,
the fact that Metaphysical Realism really is assumed in deriving the
Bell inequalities, and so, in principle, one could react to the
empirical violation of the inequalities either by rejecting Locality
(and maintaining Metaphysical Realism) or by rejecting Metaphysical
Realism.  

But there is a crucial point that speaks against this interpretation.
Notice that the last sentence of the previous paragraph did not
include the perhaps-expected parenthetical ``and maintaining
Locality''.  This was because the choice beween rejecting Locality and 
rejecting Metaphysical Realism is not a choice in the ordinary sense
-- in particular, one cannot ``save Locality'' by rejecting
Metaphysical Realism.  And this is because the very idea of ``Locality''
\emph{already presupposes} Metaphysical Realism, a point that is undeniable
once we remember what we are using the term ``Locality'' to mean:  the
requirement that all causal influences between spatially-separated
physical objects propagate sub-luminally.

The point here is this:  to reject Metaphysical Realism is precisely
to hold that there \emph{is no external physical world}.  And once one rejects
the existence of a physical world, there simply is no further issue
about whether or not causal influences in it propagate exclusively
slower than the speed of light (as required by Locality).  Or put it
this way:  ``Locality'' is the requirement that relativity's
description of the fundamental structure of space-time is correct.
But relativity theory is
thoroughly ``realist'' in the sense of Metaphysical Realism.  If there
is no physical world external to my consciousness, then, in
particular, there is no space-time whose structure might correspond to
the relativistic description -- and so that description's status would
be the same as, for example, that of claims about the viscosity of phlogiston
or theories about the causes of cancer in 
unicorns:  false in the strongest possible sense.
And so the idea of giving up
Metaphysical Realism as an \emph{alternative} to giving up Locality
(relativity's account of space-time structure) is simply
nonsense.\footnote{A 
similar point is made by Raymond Chiao and John Garrison
  in Ref. \cite{chiao}.  Note also that the position being
  argued against here (that we might save Locality by rejecting
  Metaphysical Realism) commits what philosopher Ayn Rand referred to as
  ``the fallacy of the stolen concept'' -- the fallacy consisting in
  the attempt to maintain a given concept (here, `Locality') while
  rejecting a deeper concept on which the first hierarchically depends.
  The former concept is ``stolen'' because, having denied the
  underlying context which provides its meaning, one has no cognitive
  right to its use.  For a detailed discussion see
  Ref. \cite{peikoff}.}

We may put this point in formal logical terms with the assertion that
\begin{quote}
Locality $\rightarrow$ Metaphysical Realism, 
\end{quote}
the idea being that a proper fleshing-out of the meaning of
``Locality'' manifests a tacit assumption of Metaphysical Realism,
such that any meaningful talk about Locality (such as saying that it
is true) requires that 
Metaphysical Realism is already accepted as true.  

And this suggests that, if the `realism' in `local realism' is indeed
Metaphysical Realism, the conversational implication noted by Maudlin
-- that we might save Locality by rejecting Realism -- is
patently false.  We \emph{cannot} choose between rejecting Locality and 
rejecting Metaphysical Realism (with the other being ``saved'').  We
may reject Locality (and save Metaphysical Realism) -- or we may reject
Metaphysical Realism \emph{and with it any meaningful claims about
  Locality, the causal structure of the world, and literally
  everything else that every concept and theory in the entire history
  of physics has purported to be about.}  Faced with
Bell's Theorem and the empirical data showing violations of Bell's
inequalities, we must reject Locality -- or turn solipsist, i.e.,
simply shut down
cognitively and refrain from saying anything about anything.  

And so it really doesn't make any sense after all to interpret the
`realism' in `local realism' as meaning Metaphysical Realism.  At
best, the phrase would then be a pointless redundancy, much
streamlined by replacing it simply with `Locality'.  

Unfortunately, some otherwise-serious physicists do apparently
endorse a rejection of Metaphysical Realism.  One recent
example is the paper by Matteo Smerlak and Carlo Rovelli. \cite{rovelli}  
They lobby for a ``relational'' interpretation of quantum mechanics
and an abandonment of what they refer to as ``strict Einstein
realism'' -- a doctrine that they define using Einstein's own words
(``there exists a physical reality independent of ... perception'')
and which clearly matches what we are here calling Metaphysical
Realism.  

It is interesting that 
Smerlak and Rovelli refer to Metaphysical Realism as
``\emph{strict} Einstein realism'' -- the implication being that
what they are advocating as an alternative is only some \emph{less
strict} form of realism.  But, simply put, that is not the case.
What they are advocating is the complete rejection of the most
fundamental type of realism, i.e., they are endorsing solipsism.
Smerlak and Rovelli attempt to deny this:  
``It is far from the spirit of
RQM to assume that each observer has a `solipsistic' picture of
reality, disconnected from the picture of all other observers.''
Yet, clearly, this is precisely what they \emph{do} advocate:  for
example, in their analysis of a simple EPR correlation experiment,
it emerges that, when Alice and Bob get together later to compare
results, Alice need not hear Bob reporting the same value for the outcome
of his experiment that Bob himself believes he saw.  If this isn't an
example of each observer's picture of reality being disconnected from
that of other observers, it's hard to imagine what would be.

The authors
apologize for this by noting that, at least, ``everybody hears
everybody else stating that they see the same elephant he sees'' and
report that ``[t]his, after all, is the best definition of
objectivity.''   Well, perhaps it is the best definition of
objectivity that remains possible once one has abandoned
Metaphysical Realism, but it is certainly not what scientists
normally mean by ``objectivity''.

What's ``relational'' in ``relational QM'' (RQM) is reality 
itself:  there is no such thing as reality \emph{simpliciter}; there
is only reality-for-X (where X is some physical system or
conscious observer).  Advocates of RQM thus use the word ``reality''
to mean what people normally mean by the word ``belief''.  That
some fact is, say, ``real-for-Alice'' simply means (translating from
RQM back to normal English) that Alice believes it.  And,
crucially, what is real-for-Alice need not be real-for-Bob:
``different observers can give different accounts of the same sequence
of events.'' \cite{rovelli}  

This bizarre attempt at making sense of quantum theory is related to a
wider program that might be called the ``Information Interpretation of
QM''.  According to this view, the various interpretive paradoxes and
allegedly-only-apparent non-locality are explained away by
interpreting the quantum formalism to be fundamentally \emph{about}
``information''.  The quantum mechanical wave function in particular
is regarded, not as a direct description (complete or otherwise) of
physical states, but as an encapsulation of some observer's
knowledge.  It is then not so surprising that different observers
could attribute different quantum states to the same one physical
system.  Unfortunately, bringing in the concept of ``information''
raises more questions than it resolves.  For example, 
``\emph{Information?}  \emph{Whose} information?
Information about \emph{what}?'' \cite[page 215]{bell}

Indeed, the idea of interpreting quantum mechanical wave functions as
merely summarizing some observer's limited information (and not providing a
complete description of the real, external 
physical states of systems) is not some
radical new answer to the EPR ``paradox''(as suggested by Smerlak and
Rovelli).   For it was the very
\emph{point} of the EPR paper to suggest this!

Of course, Einstein and his collaborators took for granted
Metaphysical Realism, and so to them if quantum mechanics didn't
provide complete descriptions of the real states of physical systems,
that only spoke to the need to find a theory that did.  The innovation
of Rovelli and Smerlak is thus evidently to point out that this whole
line of reasoning falls apart if one rejects Metaphysical Realism.
And indeed it does, but this can hardly be considered a resolution of
any interpretive paradox, much less a refutation of the claim that
Bell's theorem proves the inevitability of non-locality.  For Smerlak
and Rovelli's theory (which they claim as ``local'') emerges, on
inspection, to be local only in an empty sense (the only sense
possible in the context of solipsism):  \emph{everything that happens,
  happens in the same place -- namely, inside my
  head.}\footnote{Though 
technically, the concept of ``head'' (and for that matter
``inside'' and ``my'') is meaningless in the context of RQM, for, at
least as normally understood, such words refer to physical objects
(and relations between them) that exist independent of anyone's
awareness.
This is a nice illustration of the point made earlier:  once you reject
Metaphysical Realism, the whole idea of physical objects moving and
interacting in space-time -- which captures the entire content of
physics -- loses any meaning.}
There are no faster-than-light causal influences between spatially
separated physical events, simply because there \emph{are} no spatially 
separated physical events (or causal influences between them).

Tim Maudlin, in the previously cited article, remarked that 
\begin{quote}``[i]f there is something objectionable
about [accepting nonlocality and accordingly rejecting relativity
theory as the final word in space-time structure], we should consider
carefully just how objectionable it is, since there is no point in
doing something even \emph{more} objectionable just to retain the
relativistic account of space-time.''
\end{quote}
This is a fantastic argument against the Many Worlds Interpretation
considered in the previous section, with its ridiculously extravagent
ontology and its need to reject one of the fundamental philosophical 
principles underlying modern empirical science.  It's hard to imagine
how anyone could consider it reasonable to give up so much for
so (relatively) little. 
Our point here is that the corresponding argument against 
Relational/Informational Quantum Mechanics is \emph{even stronger}:
in this case, it's not just that one is giving up a lot to save a
little, but that one is giving up \emph{everything} to save
\emph{nothing}.  

In the paper in which he revealed his now-famous hoax, Alan Sokal had
this to say about the post-modern nonsense his hoax article had parodied:
\begin{quote}
``What concerns me is the proliferation, not just of nonsense
and sloppy thinking per se, but of a particular kind of nonsense and
sloppy thinking:  one that denies the existence of objective realities,
or (when challenged) admits their existence but downplays their practical
relevance.  ....  There is a real world; its properties are not merely
social constructions; facts and evidence do matter.  What sane person
would contend otherwise?''\cite{sokal}
\end{quote}
It is depressing indeed that this same kind of nonsense and sloppy
thinking is being taken seriously by some eminent physicists
as an alternative to Bell's clarity and insight into foundational
questions in quantum physics.

\section{CONCLUSIONS}

We have surveyed four different `realism' concepts.  Each has some
relation to Bell's Theorem and related issues.  Yet none of them has
provided a promising candidate for what users of the phrase `local
realism' mean by `realism' -- which leads me to speculate that the
users of that phrase don't, themselves, know what they mean, and that
the phrase has, in fact, become widespread through sheer, unthinking
inertia.  At very least, I hope the present analysis will put users of
this dubious phrase on the defensive:  anyone who claims that Bell's
Theorem is a theorem about `local realist' theories (and/or who
claims that the associated experiments have empirically refuted
`local realism' and thus leave us with a choice between rejecting
Locality and rejecting Realism) needs to explain clearly what they 
mean by `realism' and show precisely where such `realism' is assumed
in the derivation of Bell's inequalities.

How did the phrase `local realism', 
whose meaning is so unclear, appear in the first
place?  Where did it come from and why has it persisted?  I spent some
time searching the literature for this phrase, but I am by no means
confident that the earliest example I found (d'Espagnat's quoted in
the introduction) represents Patient Zero.
So I don't know for sure how to answer these questions.  But I
will offer here some speculations.

The best hypothesis I can come up with is that the phrase `local
realism' is meant to capture, simultaneously, several views held by
quantum theory's most famous critic:  Albert Einstein.  Einstein, as
the creator of relativity theory, certainly endorsed Locality (and, I
think, would clearly have endorsed Bell's mathematical formulation 
thereof).  Einstein was also a Metaphysical Realist -- a point
captured perhaps most eloquently by Wolfgang Pauli, in a 1954 letter
to Max Born, who seemed reluctant to accept that it was Metaphysical
Realism, and not an insistence on determinism, which constituted Einstein's
jumping-off point for dissatisfaction with quantum theory.  Here is
the relevant portion of the letter:
\begin{quote}
``Einstein gave me your manuscript to read; he was \emph{not at all}
annoyed with you, but only said that you were a person who will not
listen.  This agrees with the impression I have formed myself insofar
as I was unable to recognise Einstein whenever you talked about him in
either your letter or your manuscript.  It seemed to me as if you had
erected some dummy Einstein for yourself, which you then knocked down
with great pomp.  In particular, Einstein does not consider the
concept of `determinism' to be as fundamental as it is frequently held
to be (as he told me emphatically many times), and he denied
energetically that he had ever put up a postulate such as (your
letter, para. 3): `the sequence of such conditions must also be
objective and real, that is, automatic, machine-like, deterministic.'
In the same way, he \emph{disputes} that he uses as a criterion for
the admissibility of a theory the question: `Is it rigorously
deterministic?'  Einstein's point of departure is `realistic' rather than
`deterministic'...'' \cite[page 221]{bel}
\end{quote}
Or, as Einstein himself elaborated his belief in Metaphysical Realism:
\begin{quote}
``If one asks what ... is characteristic of the world of ideas of
physics, one is first of all struck by the following:  the concepts of
physics relate to a real outside world, that is, ideas are established
relating to things such as bodies, fields, etc., which claim a `real
existence' that is independent of the perceiving
subject...''\cite[170]{bel}
\end{quote}
Finally, as discussed in an earlier section, Einstein evidently
believed in deterministic non-contextual hidden variables for (at
least, it would seem) the class of experiments relevant to the 
EPR-Bell correlations.  (He believed in them because of the EPR
argument:  the only way to account for the correlations
\emph{locally} is to posit such hidden variables. \cite{nonlocchar})  
In the language of the present paper, this means that Einstein
advocated (at least in some domain) Naive Realism.  

My hypothesis is then that the contemporary phrase `local realism'
represents a kind of sloppy packaging of these three principles endorsed
by Einstein:  Metaphysical Realism, Locality, and Naive Realism.
Then, in a kind of perpetuation of the old Bohr-Einstein debates, many
contemporaries insist on seeing virually all interpretive issues surrounding
quantum theory along the following party lines:  Bohr vs. Einstein,
which gets translated into:  (orthodox) quantum mechanics vs. local
realism.  

The first part of my hypothesis is supported by the widespread
use of the phrase `local realism' to underwrite what might otherwise
be rather blatant equivocations on the term `realism'.  For example,
consider the following passage from a recent essay by Anton 
Zeilinger:
\begin{quote}
``most physicists view the experimental confirmation of the quantum 
predictions [i.e., the observed violations of Bell's inequality] 
as evidence for nonlocality.  [I don't think he's right about
``most''.  Most physicists believe this supports orthodox QM as
against ``local realism'', i.e., supports Bohr as against Einstein.
But, continuing...]  But I
think that the concept of reality itself is at stake, a view that is
supported by the Kochen-Specker paradox.  This observes that even for
single particles it is not always possible to assign definite
measurement outcomes, independently of and prior to the selection of
specific measurement apparatus in the specific experiment.''
\cite{zeilinger} 
\end{quote}
And, Zeilinger goes on to conclude, ``the distinction between reality
and our knowledge of reality, between reality and information, cannot
be made.''  And finally:  ``what can be said in a given situation must
... define ... what can exist.''
Summarizing the apparent logic: 
the Kochen-Specker theorem shows that Naive Realism 
is false.  And therefore,
Zeilinger concludes, ``the concept of reality itself'' is refuted.
There is no reality (in the sense of Metaphysical Realism) -- only
information, i.e., ideas in our minds.  ``What can be said'' defines
``what can exist.''

But, as we can now plainly see, this is simply an
equivocation.  That Naive Realism is false, doesn't entail that
Metaphysical Realism is false.  But packaging these (and more) into a
single phrase -- whose meaning is roughly ``all that stuff Einstein
believed after he went senile'' -- obfuscates any such fine distinctions.
Avoiding such equivocations
(and the ridiculously, if not viciously, extravagent conclusions to
which they lead) is the reason we must more carefully
scrutininize any use of the term `realism'.  

The other half of my hypothesis about the origins and inertia of
`local realism' is supported by the widespread belief
that the experimental tests of Bell's
inequality constitute an \emph{experimentum crucis} between orthodox
quantum theory and deterministic/realistic/hidden-variable
alternatives -- such that the Bell-inequality-violating results
provide decisive and dramatic support for orthodox quantum theory
(and, it is often suggested, provide the final empirical proof that
Bohr was right and Einstein was wrong).  Mermin, for example, writes
that ``If the
data in such an experiment are in agreement with the numerical
predictions of quantum theory, then Einstein's philosophical position
has to be wrong.'' \cite{moon}

But misunderstanding could not be more complete.  To achieve a correct
understanding, we must begin by
unpackaging the various ideas that are confusingly tied
together by `local realism.'  Starting at the beginning, does one
accept Metaphysical Realism?  If not, there is nothing more to be said
-- at least, nothing that should be of any interest to
\emph{physicists}.  Then:  does one accept Perceptual Realism?  If
not, then there is no point discussing relativity or quantum mechanics
qua scientific theories, and no possibility of discussing how best to
make sense of the empirical data collected by Aspect and others.  

With
those preliminaries out of the way, we can finally raise the question
of Locality, i.e., respect for relativity's prohibition on
superluminal causation.  A natural first question would be:  is
orthodox quantum mechanics (OQM) a local theory?  The answer is plainly
``no''.  (The collapse postulate is manifestly not Lorentz invariant,
and this postulate is crucial to the theory's ability to match
experiment.)  And so then:  Might we
construct a new theory which makes the same empirical
predictions as orthodox quantum theory, but which restores Locality?
(In other words, might we blame OQM's apparent non-locality on the
fact that it is dealing with wrong or incomplete state descriptions?)
The answer -- provided by Bell's Theorem -- turns out to be ``no''.
We are \emph{stuck} with the non-locality, which emerges as a real
fact of nature -- one which ought to be of more concern to more
physicists. 
And we are left with a freedom to decide among the various candidate
theories (all of them non-local, e.g., OQM, Bohmian Mechanics, and
GRW) using criteria that have nothing directly to do with EPR or
Bell's Theorem -- e.g., the clarity and precision with which they can be
formulated, to what extent they suffer from afflictions such as the
measurement problem, and (looking forward) to what extent they 
continue to resolve old puzzles and give rise to new insights.  

I would like to draw specific attention to the crucial historical
point at which, I think, the community's understanding first goes
significantly off the tracks:  Einstein's objections to OQM, and the
EPR argument in particular.  Too many physicists apparently fail to
grasp the EPR argument \emph{as} an \emph{argument}.  Instead, they
understand it as merely some vague expression of a philosophical
desire for `local realism', as if this whole package had simply been
asserted arbitrarily as something Einstein liked or wanted and which
OQM, to his frustration, didn't respect.  

This is nicely (i.e., clearly, i.e., painfully) exhibited in the first
two sentences of a recent experimental report in \emph{Nature}:
\begin{quote}
``Local realism is the idea that objects have definite properties
whether or not they are measured, and that measurements of these
properties are not affected by events taking place sufficiently far
away.  Einstein, Podolsky, and Rosen used these reasonable assumptions
to conclude that quantum mechanics is incomplete.'' \cite{rowe}
\end{quote}
Kudos to Rowe \emph{et al.} for making, at least, some attempt to define the
pernicious phrase `local realism.'  But I wish to call attention to
the second sentence, in particular the statement that `locality' and
`realism' (as defined in the first sentence) were \emph{assumptions}
made by EPR.  This represents exactly the confusion I just
mentioned -- specifically, the failure to grasp that EPR presented an
\emph{argument} \emph{from} Locality \emph{to} outcome-determining
hidden variables (i.e., Naive Realism). \cite{boxes}  This argument
simply must be grasped and appreciated before one can properly
understand the meaning and implications of Bell's Theorem.

So I will conclude by pleading with the physics community to revisit
these crucial foundational issues.  We must reject the thoughtless and
confused use of terminology such as `local realism' -- and
all of the misunderstandings on which this terminology rests, and
which the terminology, in turn, helps perpetuate.  Einstein and Bell
still have much to teach us about physics -- and, indeed, about
`realism' -- but before we can learn we must set aside orthodox
dogmas and allow ourselves to actually listen.

\end{document}